\pdfoutput=1
\documentclass[a4paper]{jpconf}
\usepackage{epic}
\usepackage{eepic}
\usepackage{graphics}
\usepackage{graphicx}
\graphicspath{{eps_figs/}}
\begin{document}
\title{Test of a Liquid Argon TPC in a magnetic field and investigation of high temperature superconductors
in liquid argon and nitrogen}

\author{A.~Badertscher, L.~Knecht, M.~Laffranchi, G.~Natterer, A.~Rubbia, Th.~Strauss}

\address{Institute for Particle Physics, ETH Zurich, Switzerland}

\ead{badertscher@phys.ethz.ch}

\begin{abstract}
Tests with cosmic ray muons of a small liquid argon time projection
chamber (LAr TPC) in a magnetic field of 0.55 T are described. No
effect of the magnetic field on the imaging properties were
observed. In view of a future large, magnetized LAr TPC, we
investigated the possibility to operate a high temperature
superconducting (HTS) solenoid directly in the LAr of the detector.
The critical current $I_c$ of HTS cables in an external magnetic
field was measured at liquid nitrogen and liquid argon temperatures
and a small prototype HTS solenoid was built and tested.

\end{abstract}

\section{Test of a LAr TPC in a magnetic field}

\begin{figure}

 \setlength{\unitlength}{1mm}
\begin{minipage}[c]{.45\linewidth}
    \includegraphics[width=75\unitlength]{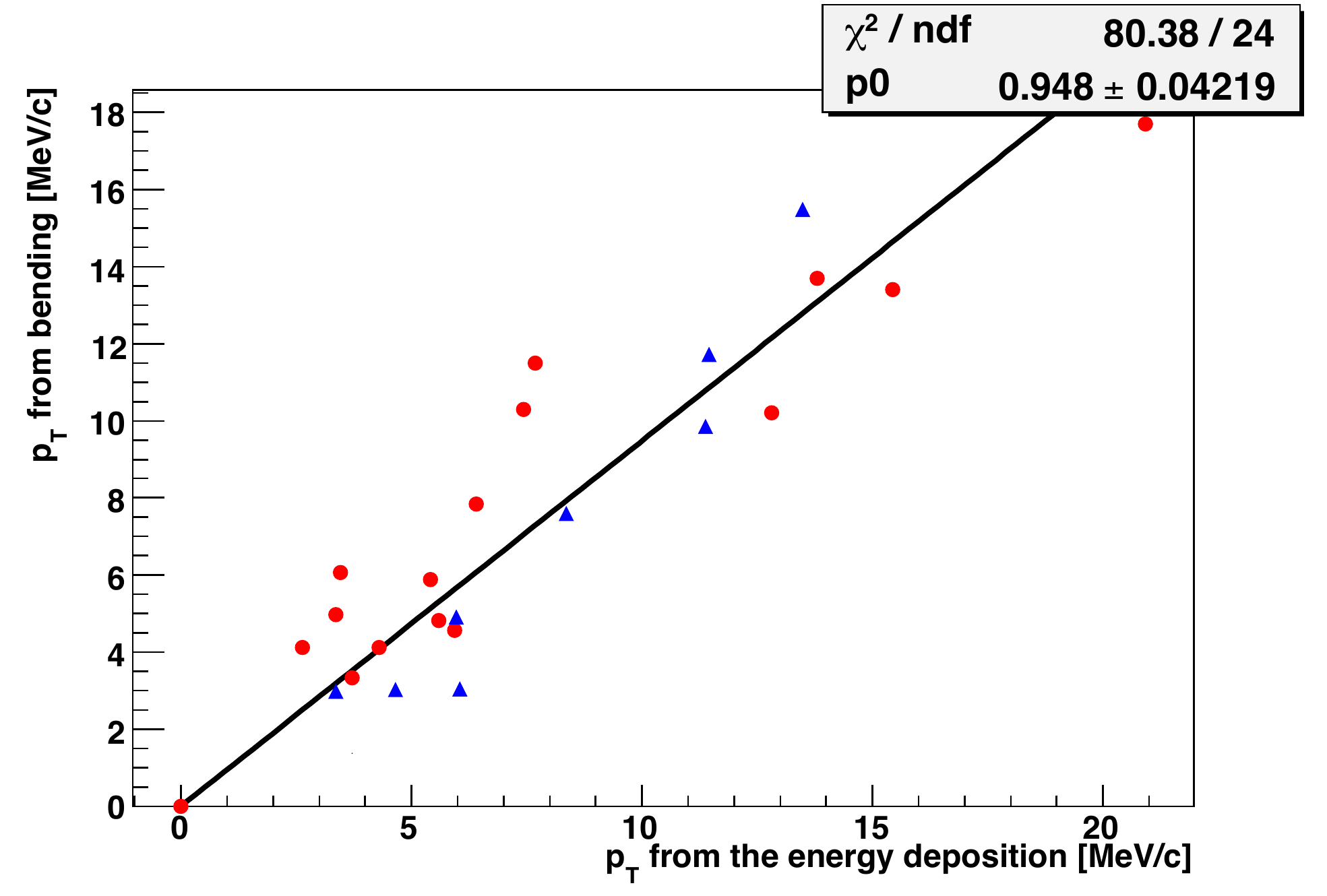}
\end{minipage}
\hspace{.05\linewidth}
\begin{minipage}[c]{.45\linewidth}
   \includegraphics[width=75\unitlength]{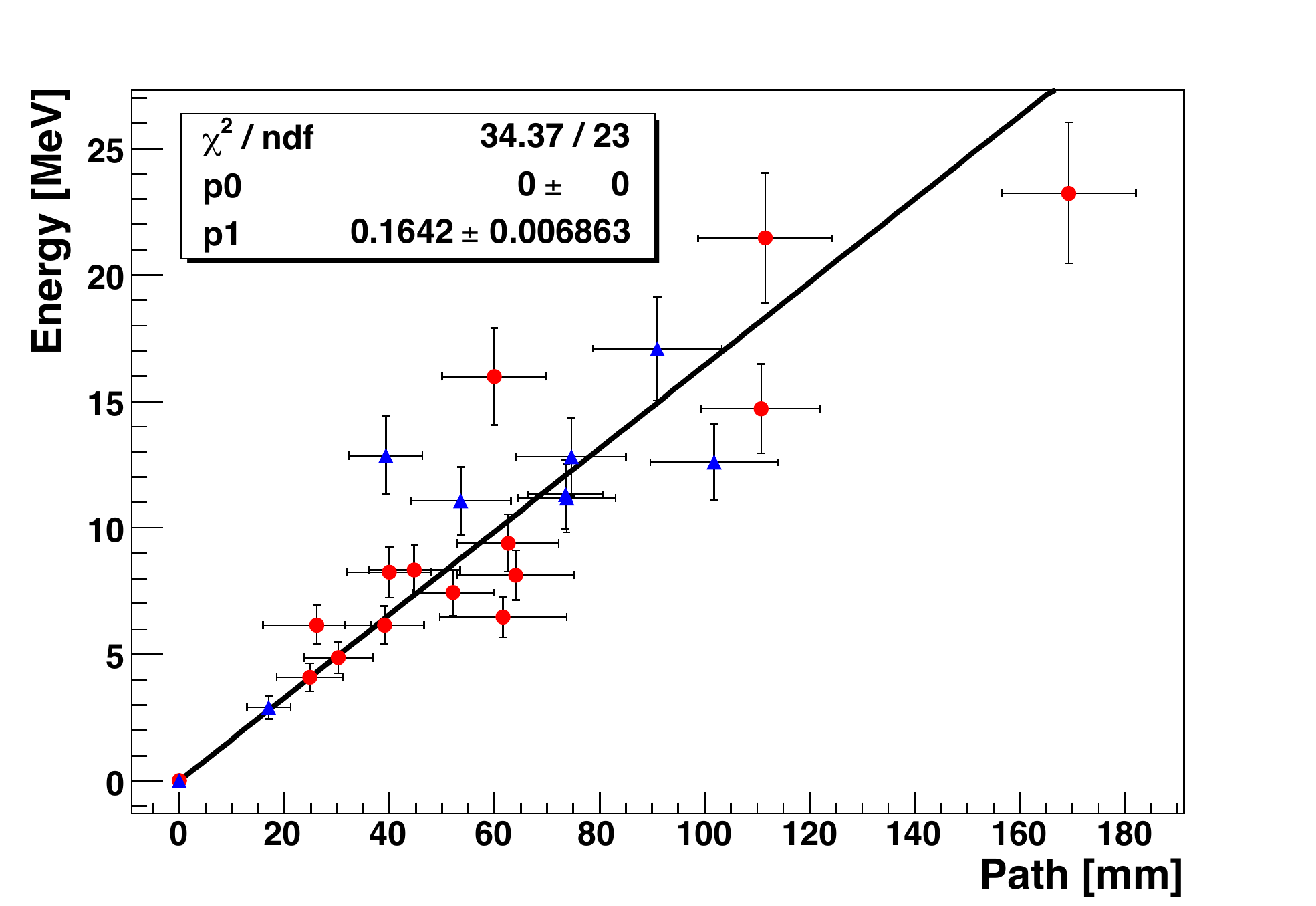}
\end{minipage}

 \caption{\label{results} left: Correlation between the two momentum measurement methods
for $\delta$-electrons (circles) and decay-positrons (triangles);
right: Measured energy as a function of the reconstructed path
length of the $\delta$-electrons (circles) and the decay-positrons
(triangles).}

 \end{figure}

The liquid argon time projection chamber (LAr TPC) \cite{Rubbia77}
is a homogeneous 3D tracking device for charged particles with
excellent - bubble chamber like - imaging properties and at the same
time it is a fine grain calorimeter for fully contained particles
due to the measurement of the energy loss dE/dx. The ICARUS
collaboration has demonstrated the feasibility of this novel
technology for large mass detectors. A 600 ton (T600) detector
consisting of two identical 300 ton half-modules was built and
successfully tested \cite{icarus04}. The detector is now installed a
the Gran Sasso underground laboratory in Italy. The possibility to
operate the LAr TPC in a magnetic field would add the very
interesting features of determining the electric charge of particles
and the momentum, also for particles leaving the chamber
\cite{Campanelli, Bueno:2001, Rubbia:2001, Bueno:2000}. The
measurement of the charge, e.g., is a must for future experiments
trying to measure CP violation in the leptonic sector at
a neutrino factory. \\
 \begin{figure}
 \begin{center}
    \includegraphics[width=8cm]{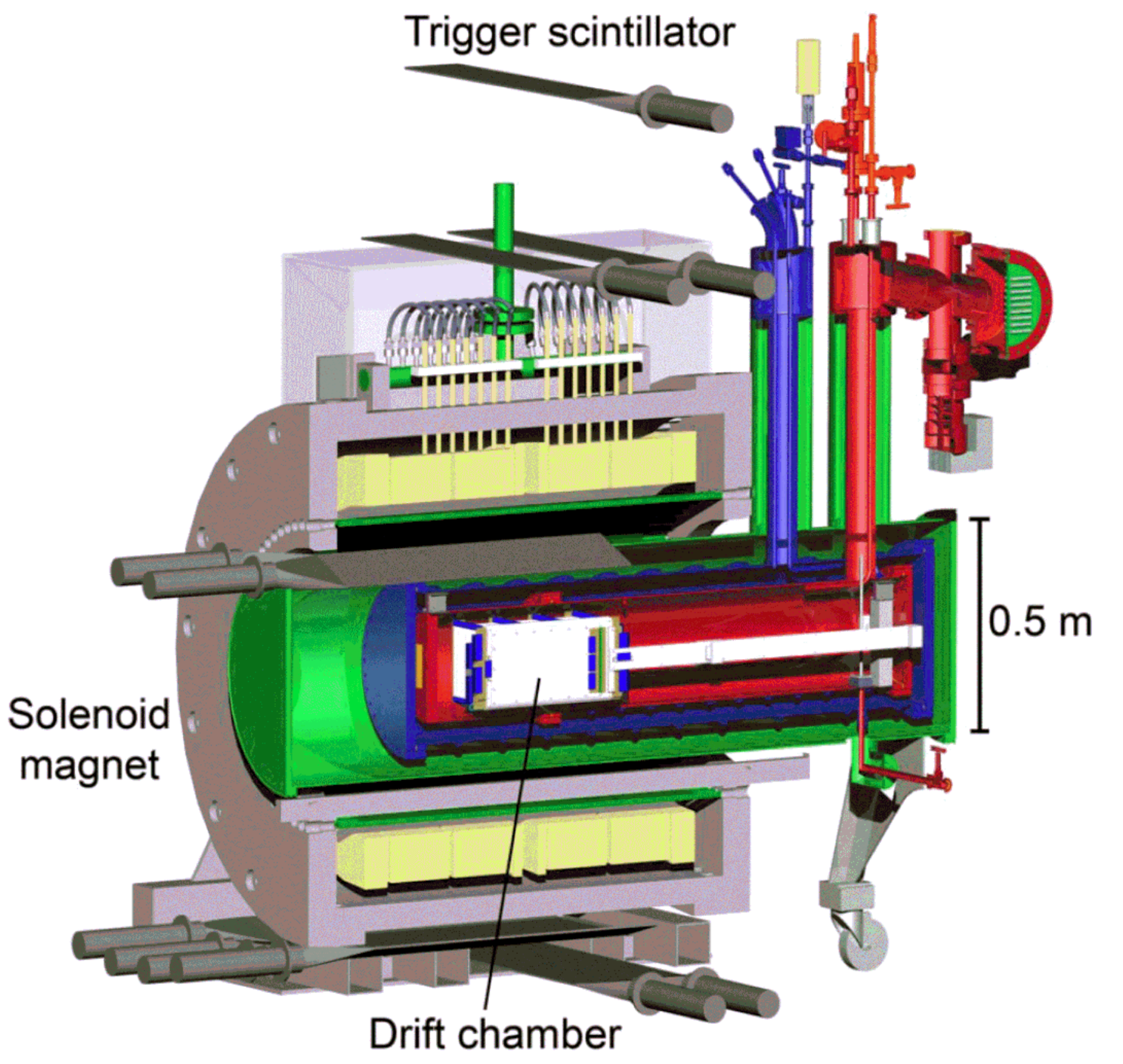}
  \caption{\label{setup} Global view of the experiment.}
 \end{center}
 \end{figure}
An R\&D program to investigate a small LAr TPC in a magnetic field
was performed. The goal was to study the drift properties of free
electrons in LAr in the presence of a magnetic field and to prove
that the imaging capabilities are not affected. A detailed
description of the experiment is given in \cite{diss_marco} and
results were published in \cite{NJP, mag_NIM}. Fig.\,\ref{setup} is
a 3D CAD drawing showing a cut through the setup with the essential
components of the experiment. The scintillators on top of the
magnet, in the bore hole on top of the cryostat, and at the bottom
of the magnet were used to trigger on cosmic ray muons. The LAr
cryostat was inserted into the recycled SINDRUM I magnet from
PSI\footnote{Paul Scherrer Institute, CH-5232 Villigen, Switzerland}
which allowed to test the chamber in a maximal field of 0.55\,T. The
cryostat consists of three concentrical stainless steel cylinders:
the innermost cylinder contains the purified LAr with the drift
chamber, the second cylinder is a LN$_2$ bath kept at a pressure of
2.7 bar in order not to freeze out the LAr at
about 1 bar, and the outermost cylinder is for the insulation vacuum. \\
\begin{figure}[b]

 \setlength{\unitlength}{1mm}
\begin{minipage}[c]{.45\linewidth}
    \includegraphics[width=70\unitlength]{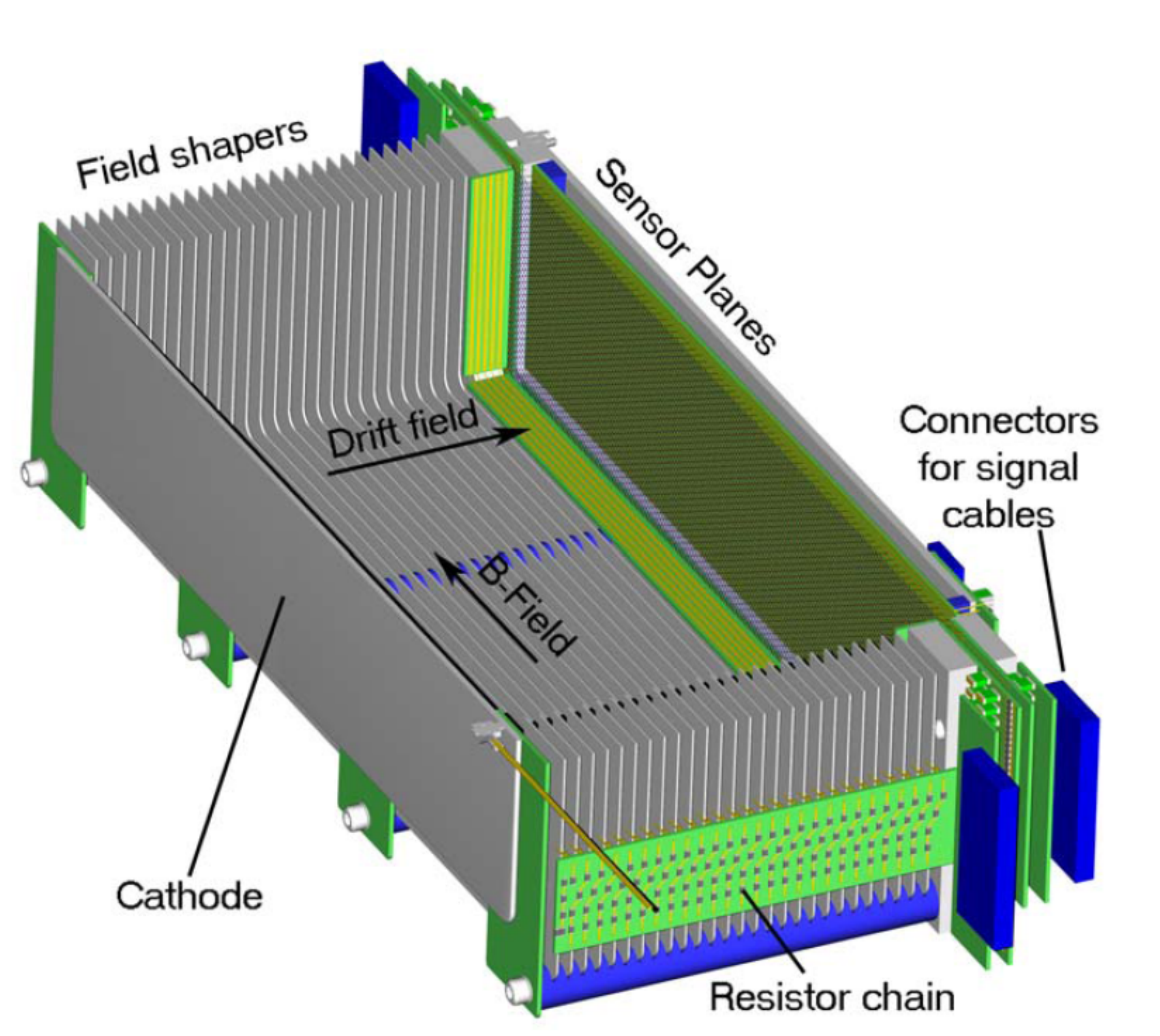}
    \end{minipage}
\hspace{.05\linewidth}
\begin{minipage}[c]{.45\linewidth}
   \includegraphics[width=60\unitlength]{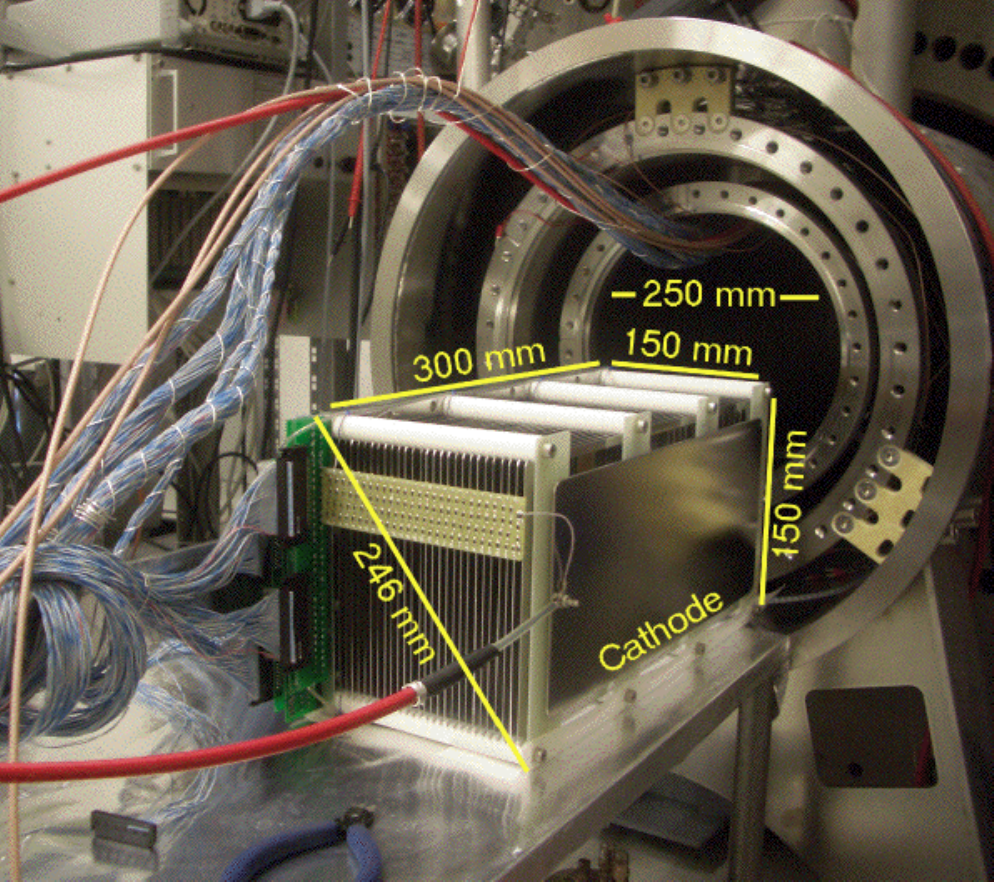}
    \end{minipage}
 \caption{\label{TPC} Left: CAD drawing of the open TPC. Right: Picture of
 the TPC ready to slide into the cryostat.}

 \end{figure}
The chamber has a length (along the B-field direction) of 300\,mm, a
height of 150~mm and a maximal drift length of 150~mm (horizontal
drift field perpendicular to the B-field). The left side of
Fig.\,\ref{TPC} shows a CAD drawing into the open chamber, and and
the right side is a picture of the chamber ready to slide into the
cryostat. After pumping the LAr cryostat, it was filled through a
purification cartridge containing activated $Cu$ powder to remove
impurities, mainly $O_2$, the LAr was not recirculated anymore
through the cartridge after the filling. \\
The chamber consists of a stainless steel cathode, 27 field shaping
electrodes to produce a homogeneous drift field and 3 sensor planes.
The first two detector planes (induction planes) are wire chambers
with $100~\mu m$ wires oriented at $\pm 60^0$ to the vertical and a
pitch of 2 mm, and the third plane (collection plane) is a PCB with
horizontal strips with a width of 1 mm and a pitch of 2 mm; the
chamber has a total of 329 channels. In the beginning of the run a
maximal drift field of 1.5\,kV/cm was applied and had to be reduced
later to to 0.3\,kV/cm because of HV breakdowns. The Lorentz angle
for a drift field of 0.5\,kV/cm and a B-field of 0.5\,T was
estimated to be $\approx 1.7^{\circ}$. \\
For initial tests without magnetic field, a coincidence of the
scintillators was used to trigger on through-going muons (trigger
rate 0.55 Hz). To trigger on stopping muons, the scintillators on
top of the cryostat in the bore hole were used, yielding the time
$t_0$ of the event needed to determine the drift time, together with
the analog sum of 32 channels of the TPC; the rate was about 1/min.
About 30'400 events were collected and visually scanned, and a small
sample of 15 $\delta$-electrons and 8 decay positrons from stopped
muons were selected for a first analysis \cite{dipl_amueller}. This
represents a highly biased sample of well measurable events fully
contained in the chamber. The tracks were reconstructed in 3D and
their momentum and kinetic energy were calculated from the magnetic
bending and the summed energy loss along the track.
Fig.\,\ref{results} shows on the left side the comparison of the
$p_t$ obtained from the magnetic bending and from the energy
measurement; on the right side is the measured kinetic energy
plotted versus the total track length.

\section{Investigation of high temperature superconductors
in liquid argon and nitrogen}

A very interesting option to magnetize a LAr TPC would be to build a
high temperature super\-conducting (HTS) solenoid directly in the
LAr cryostat of the detector. Since the manufacturers deliver data
on the temperature dependence of the properties of their commercial
HTS cables only up to LN$_2$ temperature ($77~K$), we performed a
small R\&D program to investigate the performance of different HTS
cables also at LAr temperature ($87~K$) \cite{thomas}. We tested
BSCCO (first generation HTS cable) and YBCO (second generation) from
American Superconductors AMSC\footnote{www.amsc.com} and SuperPower
Inc.\footnote{www.superpower-inc.com}. Table \ref{hts_prop}
summarizes some properties of the tested cables.
\begin{table}[h]
\begin{center}
\begin{tabular}{c|c|c|c}
AMSC  & Width  & Thickness & Critical temp. \\ \hline BSCCO & 4~mm &
0.4~mm    & 110 K \\ \hline
YBCO  & 4.3~mm & 0.25~mm    & 90 K \\
\hline \hline
SuperPower & & & \\
\hline
YBCO       & 4~mm & 0.1~mm & 90 K \\
\hline
\end{tabular}
\caption{\label{hts_prop} Properties of the tested HTS cables.}
\end{center}
\end{table}

\begin{figure}[h]
 \begin{center}
    \includegraphics[width=14cm]{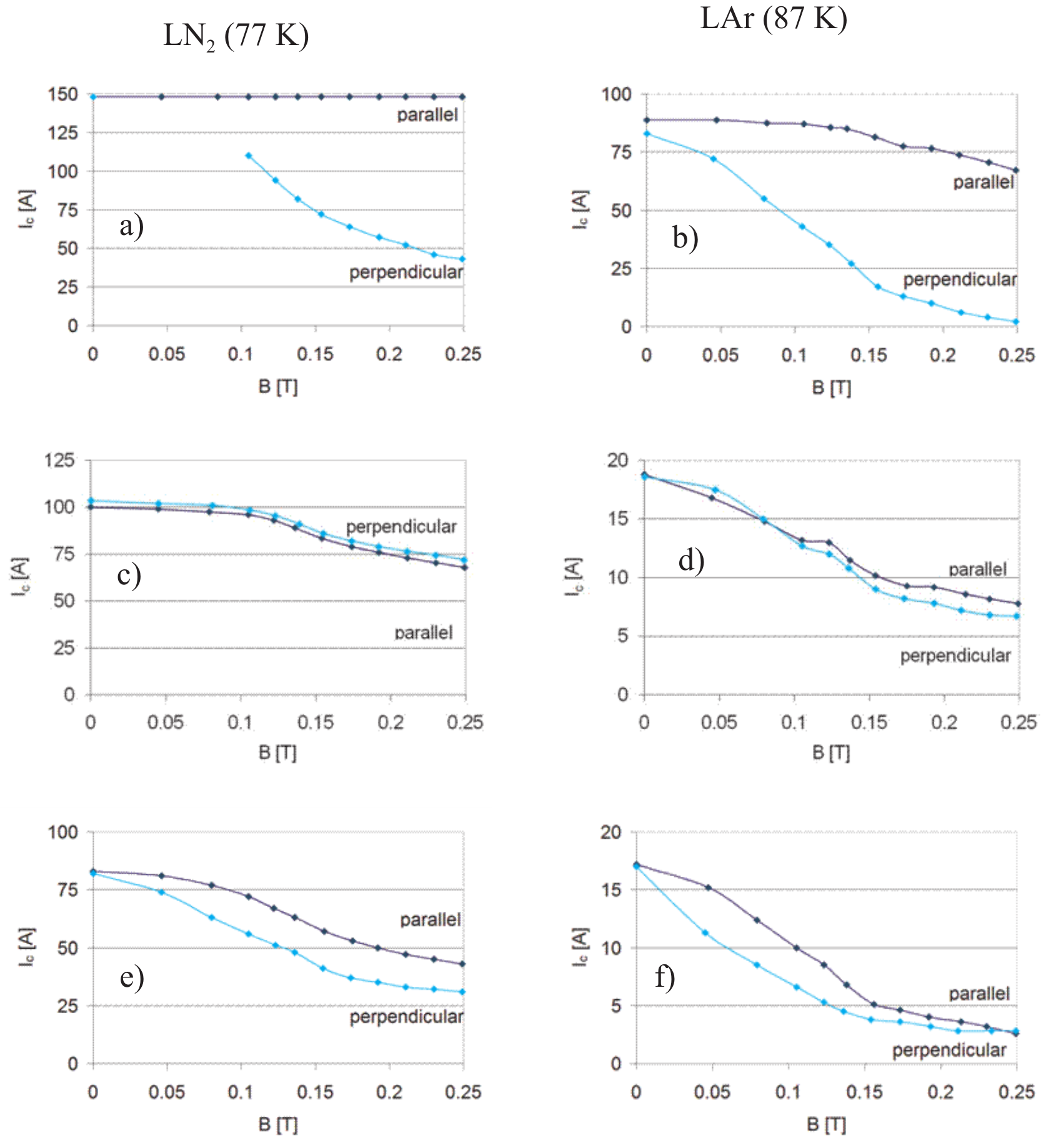}
  \caption{\label{plots} Measured critical current $I_c$ for the three tested HTS cables
  as a function of the applied parallel and perpendicular magnetic fields.
  Left column: measurements in LN$_2$ at 77\,K, right column: measurements in LAr at 87 K.
  a) and b) BSCCO cable from AMSC, c) and d) YBCO cable from AMSC, e) and f) YBCO cable from SuperPower Inc.}
 \end{center}
 \end{figure}

We measured the critical current $I_c$ in LN$_2$ and LAr as a
function of an external magnetic field applied parallel (in the
direction of the width of the cable) and perpendicular to the HTS
cable. Here, the critical current $I_c$ is defined as the current,
for which the voltage drop in the cable reaches 1\,$\mu$\,V/cm.
Fig.\,\ref{plots} shows, that in LAr the critical current is already
substantially reduced compared to the value measured in LN$_2$, in
particular for the second generation YBCO cables.

\begin{figure}[h]
\begin{center}
    \includegraphics[width=14cm]{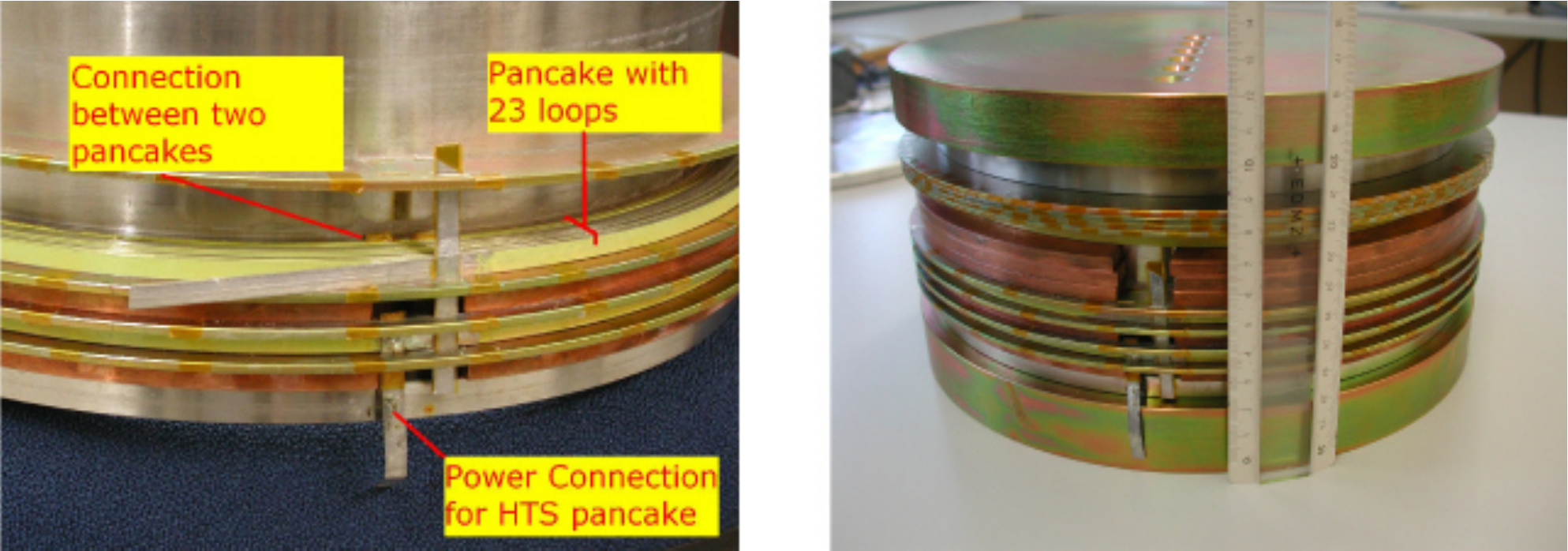}
  \caption{\label{solenoid} Left: Connections of the pancakes of the HTS solenoid.
  Right: The complete solenoid.}
 \end{center}
 \end{figure}

After testing the cables, we built a small prototype HTS solenoid
with about 100 m of the AMSC BSCCO cable \cite{diploma_thomas}. The
solenoid shown in Fig.\,\ref{solenoid} consists of four so-called
pancakes with 25 (on average) windings. It has a total length of
130\,mm, including the iron end plates and copper spacers and iron
shielding rings between the pancakes. The average diameter of the
windings was 230\,mm and the bore hole was 210\,mm. The iron
shielding rings was inserted between the pancakes to suppress the
perpendicular B-field component at the innermost coil windings. In
LN$_2$ a maximal B-field of 0.2 T was achieved at I = 145 A and in
LAr it was 0.11\,T at I = 80 A.

\end{document}